 \documentclass{aa}  
 \usepackage{graphicx}
 \usepackage{txfonts}
\begin{document}

 \newcommand{\vdag}{(v)^\dagger}
 \newcommand{\oneskip}{\vskip\baselineskip}
 \newcommand{\cc}{\hbox{$\mu$}}
 \newcommand{\ergsec}{\hbox{erg s$^{-1}$}}
 \newcommand{\aaa}{\hbox{M$_{\odot}$}}
 \newcommand{\bb}{\hbox{$\nu$}}
 \newcommand{\dd}{Optical counterpart to SAX J2103.5+4545}
 \newcommand{\ee}{SAX J2103.5+4545}
 \newcommand{\dy}{\hbox{d$^{-1}$}}

 \def\hr{\hbox{$^{\rm h}$}}                 
 \def\fhr{\hbox{$.\!\!^{\rm h}$}}           
 \def\deg{\hbox{$^\circ$}}                  
 \def\fdeg{\hbox{$.\!\!^\circ$}}            
 \def\sec{\hbox{$^{\rm s}$}}                
 \def\fsec{\hbox{$.\!\!^{\rm s}$}}          
 \def\arcm{\hbox{$^\prime$}}                
 \def\farcm{\hbox{$.\mkern-4mu^\prime$}}    
 \def\arcs{\hbox{$^{\prime\prime}$}}        
 \def\farcs{\hbox{$.\!\!^{\prime\prime}$}}  
 \def\fday{\hbox{$.\!\!^{\rm d}$}}          
 \def\per{\hbox{$^{\rm p}$}}                
 \def\fper{\hbox{$.\!\!^{\rm p}$}}          
 \def\mm{\hbox{$^{\rm m}$}}                 
 \def\fmm{\hbox{$.\!\!^{\rm m}$}}           


 \title{Optical and X-ray Outbursts of Be/X-ray binary system \\ \ee}
 \authorrunning{\"U. K{\i}z{\i}lo\u{g}lu et al.}
 \author{\"U. K{\i}z{\i}lo\u{g}lu,
        S. \"Ozbilgen,
        N. K{\i}z{\i}lo\u{g}lu,
        A. Baykal
        } 


  \institute{Physics Department, Middle East Technical University, Ankara 06531, Turkey
 }

 \date{Received -- / Accepted --}

\abstract
 {}
{  The main  goal of this study is to investigate the relationship 
between the optical and X-ray behaviours of the 
Be/X-ray binary system \object{\ee}~.  
   } 
{  We present the relations between the H$\alpha$ equivalent width, 
optical brightness and X-ray flux of the system, 
by analysing the optical photometric
and spectroscopic observations together with the X-ray observations~.
    }
{The X-ray outburst of the system occurred just after the optical outburst.
The nearly symmetric H$\alpha$ emission line profiles observed during the
beginning of the optical outburst turn into asymmetric profiles with increased EW 
values during the dissipation of the Be disc.
 A correlation between the H$\alpha$ emission line strength
and the X-ray flux was found. The H$\alpha$ line indicates the existence of
an equatorial  disc around the Be star. The 
H$\alpha$  line changed from emission to absorption during the observation 
period.
The observed double peaked HeI emission lines might come from the accretion
disc of neutron star which is temporarily formed at the time of X-ray outburst.
    }
 {}


 \maketitle

\section{Introduction}

\subsection{Be/X-ray systems}

The Be/X-ray systems consist of a Be star and a neutron star.
Be stars are close to the main-sequence. They exhibit Balmer lines in
emission and show strong infrared excess (compared 
to normal stars of the same stellar spectral type) in their spectra.
The disappearance and appearance of emission lines suggest a disc 
structure around the Be star (Slettebak 1988; Okazaki $\&$ Negueruela 2001).
The disc formation is related to the material lost from the rapidly 
rotating Be star (Porter $\&$ Rivinius 2003).
The motions of discs are rotationally dominated and quasi Keplerian 
(Hanuschik 1996). 
The interaction of neutron star with the Be disc 
(Stella et al. 1986; Negueruela et al. 1998) causes X-ray outbursts
either at the periastron passage (type I outburst with X-ray luminosity
$\sim  10^{36}$ - $10^{37}$ erg/s) or at any orbital phase (type II
outburst with X-ray luminosity greater than $10^{37}$ erg/s).
There is also X -ray quiescence phase where the accretion onto the NS
is partially or completely stopped.

In some Be/X-ray systems the Be disc forms, grows, causes an X-ray
outburst and disappears (Negueruela et al. 2001;
Reig et al. 2001; Reig et al. 2005b; Baykal et al. 2005; Reig et al. 2007).
 Some systems display an X-ray outburst and 
do not go  to  a disc loss phase (Wilson et al. 2002; Wilson et al. 2005;
K{\i}z{\i}lo\u{g}lu et al. 2007; Baykal et al. 2008). 
It is also possible to observe long periods of X-ray quiscence
even when a Be disc is present (K{\i}z{\i}lo\u{g}lu et al. 2007). 

The favoured disc model is the viscous disc model
(Okazaki 2001; Okazaki $\&$ Negueruela 2001).
Okazaki $\&$ Negueruela (2001) applied their resonantly truncated viscous
disc model to some Be/X-ray systems. They found that systems with high orbital
eccentricity show regular type I outbursts while systems with low orbital 
eccentricity display occasional giant type II outbursts.
Be/X-ray systems with
moderate eccentricity and relatively close orbits have efficient disc
truncation which allows the material to be stored in the Be disc.
When the disc radius is larger than the Roche lobe radius, type I outbursts result.
 If the Be disc is smaller than the critical lobe
radius, type II outbursts are displayed and temporary type I outbursts are
seen when the Be disc is strongly disturbed.
If the Be disc is elongated toward the periastron then material is
transferred around the periastron forming an accretion disc around the NS.

\subsection{\object{\ee}}

The transient X-ray source \object{\ee} ~was discovered on February 1997 by the {\it BeppoSAX}
Wide Field Camera (Hulleman et al. 1998). The source was active between
February and September 1997, with a pulse period of 358.61 s. The 
next activity was two years later on November 1999 and was detected by the
all-sky monitor (ASM) on board of {\it RXTE}. Using X-ray data from {\it RXTE}, 
Baykal et al. (2000) determined the orbital parameters of the system 
by Doppler shift analysis of the pulsations. These observations 
indicated a high mass companion in the system.
\object{\ee} ~had another X-ray activity in July 2002.
Orbital solutions for \object{\ee} ~showed that 
this system had a moderately eccentric orbit (e=0.4) with an orbital period of
12.66 d (Baykal et al. (2007)). Baykal et al. (2007)
 obtained a distance of 4.5$\pm$0.5 kpc to this
source. They reported a correlation between spin-up/down trend and an X-ray flux
which could be explained by accretion from an accretion disc onto the NS.

Reig et al. (2004) identified a highly reddened optical counterpart
(V=14.2 mag) to \object{\ee}. They suggested a spectral type of B0Ve for the optical
companion whose distance was suggested to be 6.5$\pm$0.9 kpc.
During their spectroscopic observations, a double peak H$_\alpha$ emission
line was observed which was an indication of a disc
surrounding Be star. Less than a month later the H$_\alpha$ line was seen in
absorption (after MJD 52870). Blay et al. (2006) presented the evolution
of H$_\alpha$ line profiles between August 2003 and September 2004.
They noticed that the transition of the H$_\alpha$ line from emission to
absorption occurs in coincidence with the transition of the Be/X-ray system
from a higher to lower X-ray activity state.
Although Be star lost its disc after the August 2003, the X-ray emission was
detected due to the accretion of material on to the NS. Reig et al. (2005a)
concluded that the X-ray emission with a luminosity of 3-5 $\times 10^{35}$
erg/s
during the quiescent phase must originate in
the stellar wind. The system was in a low activity state since the end of 2003.

The first detection of the H$_\alpha$ emission since August 2003
was made by
Manousakis et al. (2007)   on MJD 54234 and 54240 where they observed
double peaked emission lines with equivalent widths of 2.5 and 1.9 \AA,
respectively, after the report of X-ray outburst from \object{\ee} 
~(Galis et al. 2007; Krimm et al. 2007). 
Galis et al. (2007) reported a hard X-ray outburst on April 27, 2007 (MJD 54218)
which was detected by {\it INTEGRAL} IBIS/ISGRI with a flux of $ 1.7 \pm 0.1\times 
10^{-9 }$ erg/cm2/s and $ 5.3 \pm 0.7\times 10^{-10 }$ erg/cm2/s
in the 20-40 keV and 40-80 keV energy band, respectively.

In this study, we present the relationship between the long-term X-ray and the optical
variability using observations from both {\it RXTE}/ASM and ROTSEIIId as well as
the spectral analysis of the Be/X-ray system \object{\ee}.
We discuss the correlated behaviours in the optical and X-ray 
wavelengths.
The behaviour of the optical variation of the Be star helps
to understand the X-ray behaviour of the system. 
We discuss the evidences in both the H$_\alpha$ and HeI profiles.

\section{Observations and data reduction}

The optical data were obtained with Robotic Optical Transient Experiment
\footnote{http://www.rotse.net}
(ROTSEIIId)  and  the Russian-Turkish 
1.5 m Telescope \footnote{http://www.tug.tubitak.gov.tr} (RTT150)
located at Bak{\i}rl{\i}tepe, Antalya, Turkey.

\subsection{Optical photometric observations}

The ROTSEIIId telescope has an aperture of 45 cm.  It is equipped with a  
2048$\times$2048 pixel CCD and has a field of view of 
 1\deg.85$\times$1\deg.85.  Its pixel scale is 3.3\arcs/pixel. 
The system operates without filters and has a wide pass-band
which peaks at 5500 \AA ~(Akerlof et al. 2003).
 \begin{table}
 \centering
 \caption{\dd ~and photometric reference stars.}
 \label{table:1}
 \begin{tabular}{@{}ccccc@{}}
 \hline\hline
  Star  & R.A.      & Decl.     &  USNO A2  \\
        & (J2000.0) & (J2000.0) &  R mag    \\
 \hline
 Optical counterpart to\\
 SAX J2103.5+4545  & 21\hr 03\mm 35\fsec7 & +45\deg 45\arcm 04\farcs0 & 14.4  \\
 Star 1 & 21\hr 03\mm 20\fsec15 & +45\deg 45\arcm 14\farcs6 & 12.1  \\
 Star 2 & 21\hr 03\mm 21\fsec99 & +45\deg 44\arcm 55\farcs5 & 13.8  \\
 Star 3 & 21\hr 03\mm 40\fsec32 & +45\deg 46\arcm 22\farcs6 & 13.0  \\
 \hline
 \end{tabular}
 \end{table}
 \begin{figure}
 \includegraphics[clip=true,scale=0.33,angle=270]{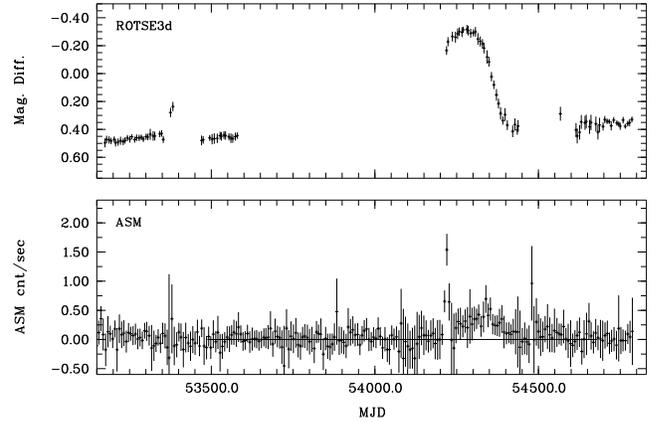}
 \caption{ Weekly averages of ROTSEIIId differential light curve of
           the Be star (top panel) and
           the X-ray light curve obtained from {\it RXTE}/ASM
           data (bottom panel).
             }
 \label{fig1}
 \end{figure}
 \begin{figure}
 \includegraphics[clip=true,scale=0.50,angle=0]{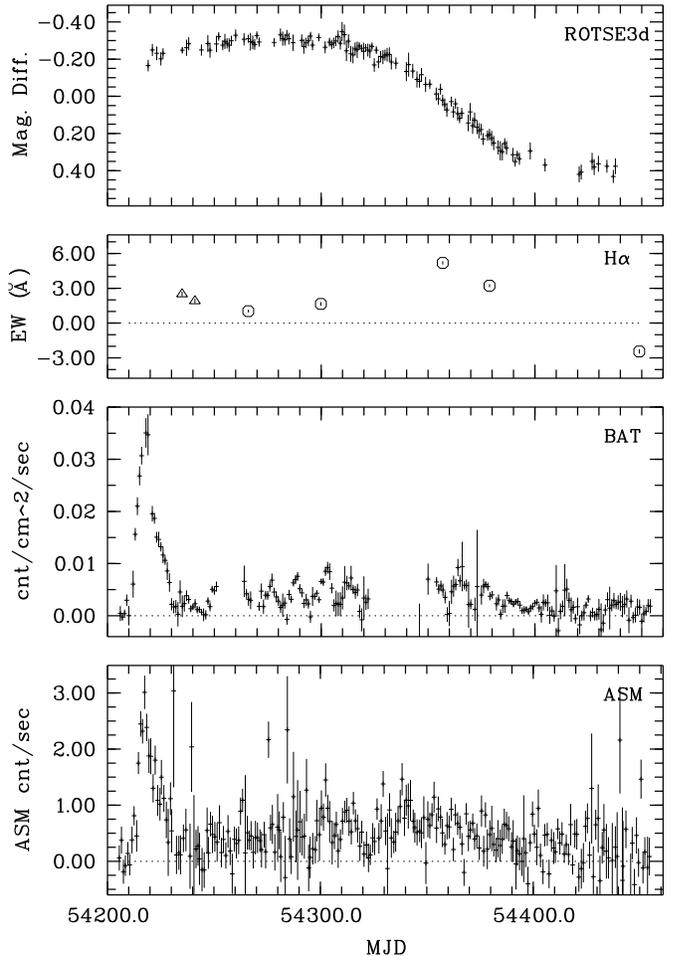}
 \caption{ Post-outburst daily averages of the differential light curve
           of the Be star (top panel)
           and the H$\alpha$ equivalent widths (second panel).
           Triangles denote values taken from
           Manousakis et al. (2007).
           The two bottom panels show daily averages of {\it SWIFT}/BAT 
           15-50 KeV
           and {\it RXTE}/ASM 5-12 KeV X-ray light curves.
          }
 \label{fig2}
 \end{figure}

The photometric observations continue since June 2004
(MJD 53167) with a break in observations between August 2005 and April 2007.
Approximately 5000 CCD frames with a 5 second exposure time were obtained.
All images were automatically dark and flat field corrected 
 by a data extraction pipeline of ROTSEIIId 
(Akerlof et al. 2003), which feeds corrected frames to the Sextractor
package (Bertin \& Arnouts 1996).
Instrumental magnitudes were obtained using Sextractor
aperture photometry on the observed CCD frames.
ROTSE magnitudes were calibrated by comparing all the field stars against
 the USNO A2.0 R-band catalog.  Apart from the data reduction pipeline, all  
CCD images were 
checked for image quality, and PSF photometry was applied
 using MIDAS and its DAOPHOT 
package (Stetson 1987,1992). 
Barycentric corrections were applied to the times of each
observation by using JPL DE200 ephemerides.  Details on the reduction of data were 
described in K{\i}z{\i}lo\u{g}lu et al. (2005).

Three reference stars were determined by examining the system's neighbourhood
for stars relatively constant magnitudes (Table 1). The average magnitude
for reference stars was calculated after the magnitudes and magnitude errors
for these stars were obtained from the appropriate frames. 
To obtain the differential magnitude, the average of the reference star's magnitudes 
is subtracted from the Be star's magnitude (Fig.1 top panel).

\subsection{Optical spectroscopic observations}
RTT150 was used for spectroscopic
observations with the medium resolution spectrometer TFOSC
(T\"UB\.ITAK Faint Object Spectrometer
and Camera) on the focal plane of the telescope.
The camera has a
2048$\times$2048, 15$\mu$
 pixel Fairchild 447BI CCD whose fov is 13$\times$13 arcmin with a pixel
scale of 0.39\arcs/pixel.
We used grism  G8 (5800-8300 \AA)
with average dispersion $\sim$1.1 \AA/pixel.
Eight spectroscopic observations have been made with TFOSC
starting in June 2007 until September 2008. A journal of optical spectroscopic
observations of \object{\ee} ~is given in Table 2. The spectra were reduced
by using MIDAS\footnote{http://www.eso.org/projects/esomidas/} and its packages:
Longslit context and ALICE.

\subsection{{\it RXTE}/ASM and {\it SWIFT}/BAT observations }

 The observations of the All Sky Monitor (ASM) on board the Rossi X-ray 
Timing Explorer ({\it RXTE})
satellite (Levine et al., 1996) and the Burst Alert Telescope ({\it BAT}) on 
board of the 
{\it SWIFT}\footnote{http://swift.gsfc.nasa.gov/docs/swift/results/}
 satellite were used to compare the optical results with the X-ray observations.
The ASM energy range is from 1.3 to 12 keV. In this study we used the 5-12
keV band observations. The BAT provides observations
in the energy range 15-50 keV. The light curves of several targets are
available in the public archive of {\it SWIFT}/BAT.

\section{Results and discussion}
 
\subsection{Optical and X-ray variability}

The ROTSEIIId light curve (weekly averages) 
of the Be/X-ray binary system \object{\ee} ~was shown in
Fig.1 together with the ASM X-ray light curve in the 5-12 keV energy band.
The precision of photometry is around 0.03 mag for the individual observation. 
The optical light curve shows a very slow increase in its brightness during
the earlier period of observations.
An optical outburst was observed around the time MJD 54200.
 Such behaviour has previously been seen in only a few Be/X-ray systems
such as  4U 0115+635 (Negueruela et al. 2001; 
Baykal et al. 2005; Reig et al. 2007),
V 0332+53 (Negueruela et al. 1999; Goranskij 2001; K{\i}z{\i}lo\u{g}lu 
et al.  2008) 
and  A 0535+26 (Clark et al. 1999; Negueruela et al. 2000; Coe et al. 2006).
The brightening of the system is of about a magnitude of 0.7. 
We infer from such a brightening that a disc is formed
 in the equatorial plane surrounding the Be star.
The Be disc is assumed to be fed from the material that is lost from the Be
star.
Fig.2. shows the plot of daily averages of the observations
during the optical and X-ray outbursts.
The same figure also shows the H$\alpha$ equivalent width (EW) values.
 During the optical outburst of the system ASM observations  show
increase in the count rates. BAT observations also show a similar picture. 
The X-ray brightening lasts more than one orbital period.
 \begin{figure}
 \includegraphics[clip=true,scale=0.33,angle=270]{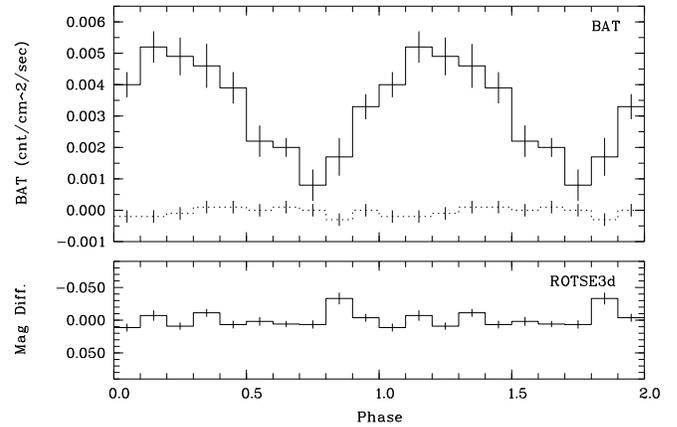}
 \caption{Top panel: The solid line shows {\it SWIFT}/BAT 15-50 KeV
           post-outburst observations folded at the period
           12.665 d using the ephemerides
           (T= MJD 52550.736+12.665N).
           The dotted line shows the pre-outburst observations. 
           Bottom panel: Folded optical light curve at the same orbital period
           after long term  variation is removed.
          }
 \label{fig3}
 \end{figure}
  \begin{figure}
 \includegraphics[clip=true,scale=0.33,angle=270]{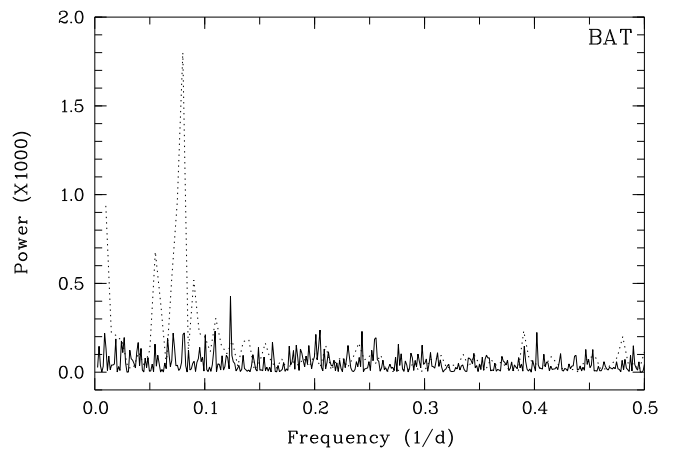}
 \caption{Power spectra of \object{\ee} ~obtained from the 15-50 keV {\it SWIFT}/BAT 
          pre-outburst (solid line) and post-outburst (dotted line)
          observations. 
          }
 \label{fig4}
 \end{figure}

According to the model of Okazaki $\&$ Negueruela (2001)  
the present system brightens in optical wavelenghts until the Be disc is
truncated by the NS. 
X-ray outburst occurred at MJD 54213 (about 3 days 
after the periastron passage, PAP),
reaching its maximum value at MJD 54218 (Krimm et al. 2007)
just before  the Be star reaches its maximum brightness.
The X-ray activity continues after the X-ray outburst and periodic small
amplitude outbursts are seen superimposed on an increased X-ray 
emission after MJD 54250. The  average  X-ray luminosity increases till
MJD 54350 and then decreases with the  brightness  of the Be star.
Considering the X-ray flux ($ 1.7 \pm 0.1\times 
10^{-9 }$ erg/cm2/s)  detected by {\it INTEGRAL} (Galis et al. 2007),
the luminosity of the source was calculated as
$\sim 4 \times  10^{36}$ and $\sim 8.5 \times  10^{36}$ erg/s
for distances 4.5 kpc (Baykal et al. 2007) and 6.5  kpc (Reig et al. 2004),
 respectively.
These luminosities are similar to the previously measured X-ray
outburst values (Baykal et al. 2007; Camero Arranz et al. 2007).
Although the luminosity is hardly $ 10^{37}$ erg/s, this outburst may be
called a type II outburst (Camero Arranz et al. 2007). As seen in the ASM and BAT
observations (Fig.2), the duration of the outburst is more than one orbital
period. After the type II outburst is completed, type I outbursts are observed
at each periastron passage of the NS. This X-ray activity lasts about
6 months till the Be disc disappears completely.
Considering a mean X-ray luminosity of $\sim 6 \times  10^{36}$ erg/s
the mass accretion rate is found to be $\sim 3 \times  10^{16}$ gr/s.
This implies a mass transfer of $\sim 3 \times  10^{-11}$ \aaa ~in 20 days
during the X-ray outburst which is a small fraction of the total mass of 
the Be disc (Reig et al. 2007).
 \begin{figure}
 \includegraphics[clip=true,scale=0.33,angle=270]{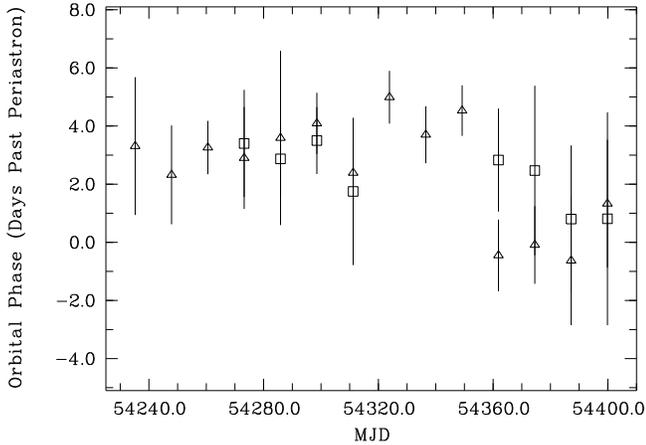}
 \caption{Orbital phase of \object{\ee} ~type I outburst peaks versus time. 
          The open squares and open triangles indicate the  
          orbital phases estimated 
          from the {\it SWIFT}/BAT 15-50 KeV
          light curve and {\it RXTE}/ASM 5-12 KeV light curve,
          respectively.
          }
 \label{fig5}
 \end{figure}

We folded the optical and
X-ray observations at the period of 12.665 d using the
epoch of the periastron passage time
 (T=MJD 52550.736+12.665N) in order to see whether 
there is any orbital period signature 
in the optical and X-ray light curve or not.
A polynomial was fitted to the optical light curve to remove the 
long term variation, and the residual time 
series were folded with the expected period. 
The folded optical and {\it SWIFT}/BAT light curves for the 
observations after MJD 54200 
are shown in Fig.3. A similar curve obtained from the
{\it SWIFT}/BAT observations prior to the X-ray outburst is also shown in
the same figure.
There is no modulation of the optical light curve at the orbital period 
 (with a statistical significance $\chi^{2}_{\nu}$ = 3.1 ).
However, the orbital period is detected 
with a $\chi^{2}_{\nu}=12.5$
for X-ray observations obtained after the X-ray outburst. 
The orbital modulation in the X-ray observations
shows the existence of type I
outbursts which occur periodically at each PAP of NS.
Pre-outburst time series does not show any  modulation.
Figure 4 shows the power spectrum of BAT observations obtained
by using the Scargle (1982) method. The calculated frequency for post
X-ray outburst observations is 0.079 d$^{-1}$ (12.6 d), which is the orbital
frequency (period) of the system.
 The detection is above 5$\sigma$ confidence level. 
We also searched for short-term variabilities in the optical light curve
over a frequency range from 0 to 20 d$^{-1}$. 
No variability is observed on shorter time scales.

The time for the  type I outburst peaks were estimated to see the orbital phasing
 of outbursts. We fitted 
a Gaussian model to the type I outbursts obtained from the daily  averaged 
light curves of 
 {\it SWIFT}/BAT (15-50 keV) and {\it RXTE}/ASM (5-12 keV) observations.
We centered the Gaussian model on the approximated
type I outburst profiles. 
In the fitting, the center value, the amplitude and the width of Gaussian
were taken as free parameters. 
We then found the peak time of the outburst.
 A similar application has been done previously 
for the Be/X-ray system EXO 2030+375 (Baykal et al. 2008). 
In Fig. 5,  we present the calculated peak time of the observed outburst 
minus the calculated time of the periastron passage at the time of the outburst
 versus time.
 The times of outburst peaks estimated from Gaussian fits to both  the 
{\it RXTE}/ASM and 
 {\it SWIFT}/BAT observations vary about 3 days after PAP. 
Baykal et al. (2000) and \.Inam et al. (2004) found a similar time delay
after PAP when the X-ray flux was maximum. 
A slowly precessing density wave in the Be disc may induce type I outbursts
at the time 3 days after the PAP time, when the Be disc is elongated toward
the periastron.
The 
larger differences after MJD 54350,
during the dispersion of the Be disc, 
between the outburst peak times for ASM and BAT observations are probably 
 due to the difficulty in determining the peak positions of the outbursts.
The X-ray outburst luminosity decreases as the Be disc disappears,
and the uncertainities in the lower energy {\it RXTE} observations are larger  
compared to those of BAT.

\subsection{Analysis of the H$\alpha$ profiles}

The observed H$\alpha$ profiles and their EW values are given in Fig.6
and Table 2, respectively. The first two double peaked emission
profiles have a similar structure and EWs  as those
given by Reig et al. (2004).
The EW values of the first two profiles  
are also similar to those of Manousakis et al. (2007)
who made observations about a month after the X-ray  outburst. 
The other two emission profiles have a more complex structure. Only a blue peak
is visible. 
 \begin{table*}
 \centering
 \caption{Journal of spectroscopic observations for the H$\alpha$ and HeI lines.}
 \label{table:2}
 \begin{tabular}{lcccccc}
 \hline
 Date & MJD 
      & EW$^{\mathrm{a}}$ (H$\alpha$)
                      & EW (HeI) & EW (HeI) \\
      & (\AA) & (\AA) & 6678 \AA & 7065 \AA \\
 \hline\hline
 2007 Jun  14 & 54265.9& 1.03$\pm$0.12 
                          &-0.32$\pm$0.11 &-0.23$\pm$0.11 \\
 2007 Jul  18 & 54299.9& 1.66$\pm$0.15 
                          &-0.49$\pm$0.10 &-0.12$\pm$0.09 \\
 2007 Sep  13 & 54356.9& 5.20$\pm$0.14 
                          & 0.56$\pm$0.11 & 0.67$\pm$0.12 \\
 2007 Oct  05 & 54378.8& 3.22$\pm$0.13 
                          & 0.59$\pm$0.14 & 0.65$\pm$0.09 \\
 2007 Dec  14 & 54448.8&-2.46$\pm$0.19 
                          &-0.75$\pm$0.36 &-0.69$\pm$0.15 \\
 2008 Jun  19 & 54636.9&-2.54$\pm$0.18 
                          &-0.61$\pm$0.16 &-0.64$\pm$0.27 \\
 2008 Aug  26 & 54704.9&-2.32$\pm$0.18 
                          &-0.93$\pm$0.32 &-0.71$\pm$0.16 \\
 2008 Sep  22 & 54731.0&-2.44$\pm$0.16 
                          &-0.75$\pm$0.27 &-0.84$\pm$0.18 \\
 \hline
 \end{tabular}
 \begin{list}{}{}
 \item[$^{\mathrm{a}}$]
      Positive values stand for emission line EWs whereas absorption 
      EWs are denoted as negative. 
 \end{list}
 \end{table*}
  \begin{figure}
 \includegraphics[clip=true,scale=0.46,angle=0]{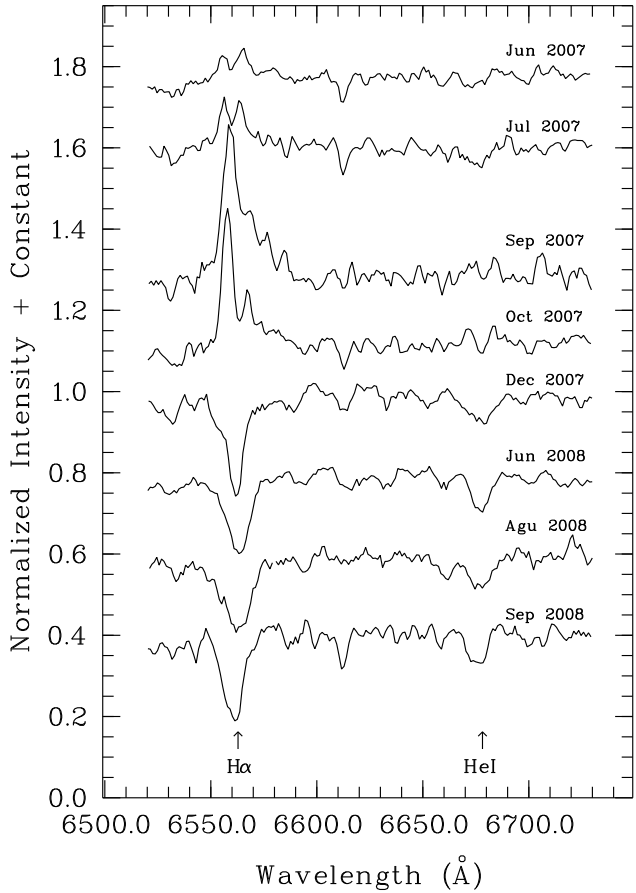}
 \caption{ H${\alpha}$ and HeI ($\lambda$ 6678) profiles observed after 
            The type II outburst.  Each profile normalised and 
            properly offsetted.
          }
 \label{fig6}
 \end{figure}
In Fig. 2 we also plot the EW values of the H$\alpha$
emission line given by Manousakis  et al. (2007)
in addition to our values.
The system shows weak H$\alpha$ emission at the time of the maximum optical
brightening of the system.  
The EWs observed about four  months after the X-ray outburst are larger
than its previous values. There seems to be no correlation between the H$\alpha$
emission EWs and the optical brightness of the system. After MJD 54400 the Be disc
disappears, and the H$\alpha$ lines are observed in absorption. Such disc
formation and disc loss phases are seen in A0535+26 (Clark et al. 1999,
Negueruela et al. 2000)
and 4U 0115+63 (Negueruela et al. 2001, Baykal et al. 2005, Reig et al. 2007).

Negueruela et al. (1998) noted that the H$\alpha$ emission region was affected
during the type II outburst by the perturbation in the Be disc which occurred
immediately after the type II outburst. When the
density perturbations interact with the NS type I
outbursts occur. They presented observational evidence for large density
perturbations in the Be discs of Be/X-ray systems like 4U 0115+634,
V0332+53 and A0535+262,  which are moderately eccentric systems.
Asymmetric shapes of the H$\alpha$ line profiles are due to density
perturbation which develops V/R variability.
The variations in the shape of the H$\alpha$ emission profiles (V/R
variability) show a global change in the structure of the Be disc.
Negueruela et al. (1998) explained the observed V/R variations with the models of one-armed 
oscillations.
They found a correlation between type II outburst and density perturbation.
Reig et al. (2005b) suggested that the effect of density perturbation was not
seen in the shape of the H$\alpha$ lines until the disc was fully formed,
that is after the disc size and density reached  a critical value.
In the present study the H$\alpha$ profiles do not reflect a perturbation
just after the X-ray outburst.
 They show a quasi symmetric form. 
At the time of the X-ray outburst it is not clearly understood whether the Be 
disc is distorted or not.
When the system reaches its maximum brightness and the Be disc is
fully developed the EWs of these profiles are 
less than $\sim$2 \AA.
The first asymmetric profile is seen when EW of the H$\alpha$ line is $\sim$5 \AA.
We observe asymmetric H$\alpha$ line profiles with sharp blue peak and
larger EW values during the decay of the Be disc. This indicates
a structural change in the outer layers of the Be disc after MJD 54300.
If a density perturbation is responsible for the complex profile
of the H$\alpha$ it is observed during the dissipation of the Be disc when
the size or density of the disc decreases.

As Reig et al. (2007) explained, the surface density of the Be disc increases
rapidly in Be/X-ray binaries due to the truncation of the disc. The disc
becomes optically thick at IR wavelengths, and becoming instable, it
begins to warp, tilt and precess.
The high values of the H$\alpha$ EW in 4U 0115+63
happen because of a warped disc which presents a larger surface area to 
the observer.
In the present system \ee,
the increase in  the H$\alpha$ EW and the distortion in the structure of
the blue peaked H$\alpha$ profile during the decrease of the optical
brightening can be seen  as a result of the warping of the disc.
An instability in the Be disc causes the outer layers to
warp, showing a larger surface area to the observer.
 \begin{figure}
 \centering
 \includegraphics[clip=true,scale=0.43,angle=0]{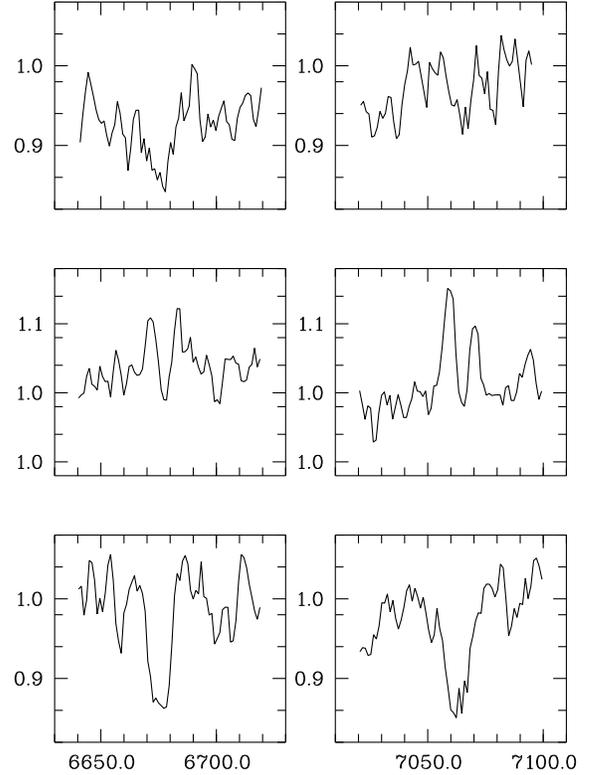}
 \caption{ HeI $\lambda\lambda$6678,7065 \AA ~profiles observed after 
            the X-ray outburst.
            Top panel: on MJD 54299.9 Mid panel: on MJD 54378.8. 
            Bottom panel: on MJD 54731.0.
          } 
 \label{fig7}
 \end{figure}

\subsection{The HeI $\lambda\lambda$6678,7065 \AA ~lines}

The EWs of the HeI $\lambda\lambda$6678,7065 \AA ~lines are  given in Table 2.
Both lines behave similarly as seen in Fig.7.
The HeI lines show a double peak emission feature in the two
spectra on MJD 54356 and 54378.
The double peaked HeI features are typical in the early Be stars without 
compact companions (Dachs et al. 1992; Negueruela et al. 2004). 
These emission lines were also observed in the Be/X-ray systems like
RX J0440.9+4431 (Reig et al. 2005b) and A0535+26 (Clark et al. 1998;
Coe et al. 2006). 
In the first two spectra obtained on MJD 54265 and 54299
the HeI lines are not seen well.
However a weak absorption 
feature can be deduced. During these observations the optical brightness 
is at its highest  value.
The HeI feature is seen in absorption in the remaining  series of the spectra 
where the optical brightness has returned to its pre-outburst value.

The  relation between the H$\alpha$ emission strength and the X-ray
flux (Fig.2) shows that the increase in the X-ray luminosity is accompanied 
by an increase in the EWs of the H$\alpha$ lines.
Higher EW values of the H$\alpha$ emission line and the HeI double peak emission lines
occur  during the decay of the Be disc.
It might be possible that  
a temporary accretion disc formed around a neutron star at the time of 
the X-ray outburst  
 contributes to the emission of the H$\alpha$ and HeI lines.
 Giovanelli et al. (2007) argued that the double
peak HeI emission lines observed in the Be/X-ray binary system
A0535+26 was formed in the temporary accretion disc of NS. They
calculated the radius for the HeI 6678 \AA ~emission line and found that 
this radius was incompatible with the stellar radius. They decided that 
the doubling
of the HeI emission lines was due to the accretion disc.
The presence of a temporary accretion disc for  \object{\ee} ~may be questionable.
We are not able to show the existence of such a temporary disc
during the last X-ray outburst. However,
Baykal et al. (2007) obtained a correlation between
spin up/down and X-ray flux for \object{\ee}
~during its previous two X-ray outbursts and explained this by
accretion from an accretion disc.
Very recently, Reig et al. (2009) found a correlation between the strength and shape of the H$\alpha$ line originated from the Be star and the X-ray emission 
from the neutron star.

\section{Summary}

Be/X-ray binary system \object{\ee} ~has been monitored since June 2004 
with a break in observations between August 2005 and April 2007. 
The system shows four X-ray outbursts
during the last ten years. The 2007 X-ray outburst coincides with the optical
outburst of the system. The optical brightening of the system of about the
magnitude of 0.7. The H$\alpha$ emission lines observed during this time show the 
existence of a disc surrounding the Be star. The H$\alpha$ emission profiles 
go from symmetric profiles to asymmetric ones in about four months. 
The appearance of HeI double peak emission lines  occurred when 
there is a structural change in the H$\alpha$ profile. Both lines 
(H$\alpha$ and HeI lines) show an increase in EW values during 
the decay of the Be disc.
The peak times of the type I outbursts are about three days after the PAP 
of the neutron star. 
Temporary type I outbursts continues about 12 orbits after the X-ray outburst.
After the decay of the Be disc  the
emission lines  of the H$\alpha$ and HeI  turn into absorption lines
 and the system is still in optic quiscence.

\section*{Acknowledgements}
We thank the anonymous referee for a careful reading and valuable comments.
This project utilizes data obtained by the Robotic Optical Transient Search
Experiment.  ROTSE is a collaboration of Lawrence Livermore National Lab,
Los Alamos National Lab, and the University of Michigan
(http://www.rotse.net).
We thank the Turkish National Observatory of T\"UB\.ITAK
for running the optical facilities.
We acknowledge support from T\"UB\.ITAK, The Scientific and Technological 
Research Council
of Turkey,  through project 106T040.
We also acknowledge the {\it RXTE}/ASM team for the X-ray monitoring data.
{\it SWIFT}/BAT transient monitor results were provided by the {\it SWIFT}/BAT team.


\begin{thebibliography}{99}
 \bibitem[2003]{Akerlof} Akerlof C. W.,
  Kehoe R. L., McKay T. A., et al. 2003, PASP, 115, 132
 \bibitem[2000]{Baykal}
  Baykal, A., Stark, M. J., \& Swank, J. H. 2000, ApJ, 544, L129
 \bibitem[2005]{Baykal}
  Baykal, A., Kiziloglu, U., Kiziloglu, N., et al. 2005, 
  A\&A, 439, 1131 	
 \bibitem[2007]{Baykal}
  Baykal, A., \.Inam, S. C., Stark, M. J., et al.
   2007, MNRAS, 374, 1108
 \bibitem[2008]{Baykal}
  Baykal, A., Kiziloglu, U., Kiziloglu, N., et al. 2008,
  A\&A, 479, 301
 \bibitem[1996]{Bertin}
  Bertin, E., \& Arnouts, S. 1996, A\&AS, 117, 393
 \bibitem[2006]{Blay}
  Blay, P., Camero, A., Martinez-Nunez, S., et al. 2006,
  (ESA SP-604), 243
 \bibitem[2007]{Camero}
  Camero Arranz, A., Wilson, C. A., Finger, M. H., Reglero, V. 2007, A\&A,
  473, 551
 \bibitem[1998]{Clark}
   Clark, J. S., Tarasov, A. E., Steele, I. A., et al. 1998, MNRAS, 294, 165
 \bibitem[1999]{Clark}
  Clark, J. S., Lyuty, V. M., Zaitseva, G. V., et al. 1999, MNRAS, 302, 167
 \bibitem[2006]{Coe}
  Coe, M. J., Reig, P., McBride, V. A., et al. 2006, MNRAS, 368, 447
 \bibitem[1992]{Dachs}
  Dachs, J., Hummel, W., \& Hanuschik, R. W. 1992, A\&AS, 95, 437
 \bibitem[2007]{Galis}
  Galis, R., Beckmann, V., Bianchin, V., et al. 2007, Atel \#1063
 \bibitem[2007]{Giovannelli}
  Giovannelli, F., Bernabei, S., Rossi, C., et al. 2007, A\&A
  475, 651
 \bibitem[2001]{Goranskij}
  Goranskij, V. 2001, AstL, 27, 516
 \bibitem[1996]{Hanuschik}
  Hanuschik, R. W., Hummel, W., Sutorius, E., et al. 1996, A\&AS, 116, 309
 \bibitem[1998]{Hulleman}
  Hulleman, F., in`t Zand, J. J. M., \& Heise, J. 1998, A\&A, 337, L25
 \bibitem[2004]{Inam}
  \.Inam, S. C., Baykal, A., Swank, J. H., et al.  2004, ApJ, 616, 453
 \bibitem[2005]{Kiziloglu}
  K{\i}z{\i}lo{\u{g}}lu, U., K{\i}z{\i}lo{\u{g}}lu, N., \& Baykal, A. 2005, 
  AJ, 130, 2766
 \bibitem[2007]{Kiziloglu}
  K{\i}z{\i}lo{\u{g}}lu, U., K{\i}z{\i}lo{\u{g}}lu, N., Baykal, A., et al. 
  2007, A\&A, 470, 1023 
 \bibitem{Kiziloglu}
  K{\i}z{\i}lo{\u{g}}lu, U., K{\i}z{\i}lo{\u{g}}lu, N., \& Baykal, A., et al.
  2008, IBVS, 5865
 \bibitem[2007]{Krimm}
  Krimm, H. A., Barthelmy, S. D., Barbier, L., et al. 2007, Atel \#1064
 \bibitem[1996]{Levine}
  Levine, A. M., Bradt, H., Cui, W., et al. 1996, ApJ, 496, L33
 \bibitem[2007]{Manousakis}
  Manousakis, A., Reig, P., \& Kougentakis, A., 2007, Atel \#1085
 \bibitem[1998]{Negueruela}
  Negueruela, I., Reig, P., Coe, M. J., et al. 1998, A\&A, 336, 251
 \bibitem[1999]{Negueruela}
  Negueruela, I., Roche, P., Fabregat, J., et al. 1999, MNRAS, 307, 695
 \bibitem[2000]{Negueruela}
  Negueruela, I., Reig, P., Finger, M. H., et al.  2000, A\&A, 356, 1003
 \bibitem[2001]{Negueruela}
  Negueruela, I., Okazaki A. T., Fabregat, J., et al. 
  2001, A\&A, 369, 117
 \bibitem[2004]{Negueruela}
  Negueruela, I., Steele, I. A., \& Bernabeu, G. 2004, AN, 325, 749
 \bibitem[2001]{Okazaki}
  Okazaki, A. T. 2001, PASJ, 53, 119
 \bibitem[2001]{Okazaki}
         Okazaki, A.T., \& Negueruela, I. 2001, A\&A, 377, 161
 \bibitem[2003]{Porter}
  Porter, J. M., \& Rivinius, T. 2003, PASP, 115, 1153
 \bibitem[2001]{Reig}
  Reig, P., Negueruela, I., Buckley, D. A. H., et al. 2001, A\&A, 367, 266
 \bibitem[2004]{Reig}
  Reig, P., Negueruela, I., Fabregat, J., et al. 2004, A\&A, 421, 673
 \bibitem[2005a]{Reig1}
  Reig, P., Negueruela, I., Papamastorakis, G., et al. 2005a, A\&A, 440, 637  
 \bibitem[2005b]{Reig2}
  Reig, P., Negueruela, I,. Fabregat, J., et al. 2005b,  A\&A, 440, 
  1079
 \bibitem[2007]{Reig3}
  Reig, P., Larionov, V., Negueruela, I., et al. 
  2007, A\&A, 462, 1081
  \bibitem[2009]{Reig4}
  Reig, P., Slowikowska, A., Zezas, A., et al. 2009, Astro-ph.HE, 0908.4497
 \bibitem[1982]{Scargle}
  Scargle, J. D. 1982, ApJ, 263, 835
 \bibitem[1988]{Slettebak}
  Slettebak, A. 1988, PASP, 100, 770
 \bibitem[1986]{Stella}
  Stella, L., White, N. E., \& Rosner, R. 1986, ApJ, 208, 669
 \bibitem[1987]{Stetson}
  Stetson, P.B. 1987, PASP, 99, 191
 \bibitem[1992]{Stetson}
  Stetson, P. B. 1992, ASP Conf. Series, 25, 297
 \bibitem[2002]{Wilson}
  Wilson, C. A., Finger, M. H., Coe, M. J., et al. 2002, ApJ, 570, 287
 \bibitem[2005]{Wilson}
  Wilson, C. A., Weisskopf, M. C., Finger, M. H., et al. 2005, ApJ, 622, 1024 
 \end{thebibliography}
\end{document}